\documentstyle[prl,aps,epsfig,floats]{revtex}

\begin{document}

\title{Evidence for incommensurate spin fluctuations in Sr$_2$RuO$_4$.}
\author{Y. ~Sidis$^a$, M.~ Braden$^{a,b}$, P. ~Bourges$^a$, B.~Hennion$^a$.
S. Nishizaki$^c$, Y. ~Maeno$^{c,d}$, Y. Mori$^c$}  
\address{$^a$Laboratoire L\'eon Brillouin,
C.E.A./C.N.R.S., F-91191-Gif-sur-Yvette CEDEX, 
France\\
$^b$ Forschungszentrum Karlsruhe, INFP, Postfach 3640, D-76021 
Karlsruhe, Germany\\
$^c$ Department of Physics, Kyoto University,
Kyoto 606-8502, Japan\\
$^d$ CREST, Japan Science and Technology Corporation, Kawaguchi,
Saitama 332-0012, Japan
}
\date{\today}


\twocolumn[\hsize\textwidth\columnwidth\hsize\csname@twocolumnfalse\endcsname
\vspace*{ -1 cm}

\maketitle

\begin{abstract}
We report first inelastic neutron scattering measurements in the normal 
state of Sr$_2$RuO$_4$ that reveal
the existence of incommensurate magnetic spin fluctuations located at
${\bf q}_0$=($\pm$0.6$\pi$/a,$\pm$0.6$\pi$/a,0). 
This finding confirms recent band structure calculations 
that have predicted incommensurate magnetic responses related to 
dynamical nesting properties of its Fermi surface.
\end{abstract}


]

\narrowtext

Being the only example
of a non-cuprate layered perovskite superconductor \cite{1},
Sr$_2$RuO$_4$ has attracted considerable attention 
despite its rather 
low critical temperature , $T_c \sim$1 K \cite{1}.
Its normal state is characterized as an essentially two-dimensional Fermi liquid and the
coherent interlayer transport settles in at low temperature only  \cite{2}.
The susceptibility is most likely dominated by an enhanced Pauli spin 
susceptibility, $\tilde{\chi}$.
Meanwhile, the Sommerfeld coefficient, $\gamma$, in the specific heat is 
enhanced by a factor 3.5 with respect to band structure calculations \cite{2,3}.
It yields a Wilson ratio, $R_W$ ($\sim \tilde{\chi}/\gamma$), of 1.7.
This value indicates that the enhancements in both susceptibility and 
electronic specific heat can be ascribed to the same origin:
most likely correlations among electrons \cite{2}.

Noticing that SrRuO$_3$ is ferromagnetic (FM), it has been conjectured that Sr$_2$RuO$_4$ 
is close to a FM instability as well \cite{4}.
This assertion is corroborated by microscopic calculation of magnetic 
properties of ruthenates \cite{5}.
Since FM fluctuations disfavor both s- and d-wave superconductivity, 
it has been suggested that superconductivity in Sr$_2$RuO$_4$ should 
possess p-wave symmetry (triplet pairing) \cite{4,6}.
Conventional local density approximation (LDA) calculations \cite{oguchi,7} 
give a correct Fermi-surface topography, probed
by de Haas-van Alphen measurements \cite{8}, as well as the magnetic
 enhancement due to Stoner exchange enhancement, although the 
mass renormalization cannot be explained within LDA calculations.
In the superconducting state, the $^{101}$Ru nuclear spin lattice relaxation rate, $1/T_1$
exhibits a sharp decrease without a  coherence peak (Hebel-Slichter peak) just above $T_c$,
 supporting the idea 
that an anisotropic pairing is effectively realized in Sr$_2$RuO$_4$ \cite{10}.
In addition, the spontaneous appearance of an internal magnetic field below the transition
temperature, reported by muon spin rotation measurements ($\mu SR$) \cite{11}, 
and the absence of $^{17}$O Knight shift modifications below $T_c$ \cite{ishida} point towards the  triplet p-wave superconductivity.
However,  few experiments have really probed the exact 
nature of the spin fluctuations.
Only the observation of a similar temperature dependence for 
$^{101}$Ru ${1/T_1T}$ and for $^{17}$O ${1/T_1T}$ in the NMR experiments
by Imai {\it et al.} \cite{12} has suggested that
spin fluctuations are predominantly FM in origin.

The determination of the antiferromagnetic order in the 
closely related compound Ca$_2$RuO$_4$ \cite{caruo,nakatsuji}
has suggested that the 
picture of a near-by FM instability in Sr$_2$RuO$_4$ is too simple.
Furthermore, recent calculations 
which take into account the particular topology of the Fermi-surface,
have predicted a sizeable magnetic response
at the incommensurate wave-vector (2$\pi$/3a,2$\pi$/3a,0) \cite{9},
i.e. far away from the zone-center.
The enhanced susceptibility arises 
from pronounced nesting properties of the almost
one-dimensional d$_{xz,yz}$ bands.
Mazin and Singh discuss the possibility of a competition
between p-wave and d-wave superconductivity in Sr$_2$RuO$_4$ \cite{9}.

In this letter, we report first inelastic neutron scattering (INS) 
measurements performed on single crystals of Sr$_2$RuO$_4$ in the normal state. 
Our data reveal dominant magnetic scattering at the incommensurate wave vectors
{\bf$q_0$}=($\pm$0.6$\pi$/a,$\pm$0.6$\pi$/a,0), i.e. very close to the 
positions 
predicted by the band-structure calculations. 
The relevance of these findings for the mechanism of superconductivity 
in Sr$_2$RuO$_4$ will be discussed.

Most of the INS measurements presented here have been carried out on a single crystal of 
cylindrical shape  (4mm in diameter and 35mm long) grown by a floating zone method.
The sample exhibits the superconducting transition at $T_c \sim$0.62 K. 
The single crystal was mounted in an aluminum can and attached to the cold 
finger of a closed cycle helium refrigerator. 
The INS experiments were performed on the triple axis spectrometers 2T (thermal beam) and 4F2 (cold beam) 
at the Laboratoire L\'eon Brillouin, Saclay, France. 
These spectrometers use neutron optics that focus the beam to the sample,
with a resulting gain of neutron flux that proved to be crucial 
for these experiments. 
The experimental set up incorporates PG002 monochromator and analyzer 
and  14.7 meV fixed final energy.
A pyrolytic graphite filter was inserted into the scattered beam 
in order to remove higher order contaminations. 
Data were taken within the scattering plane spanned by (1,0,0) and (0,1,0) directions.
Some additional measurements were performed using several smaller single crystals
with higher transition temperatures, $T_c$=1.4--1.5\ K; these experiments
have revealed similar signals.
Throughout this article, the wave vector {\bf Q}=(H,K,L) is indexed in units of
the reciprocal tetragonal lattice vectors $2\pi /a=2\pi /b=1.63$ \AA$^{-1}$ and $2\pi /c=0.49$ \AA$^{-1}$ (I4/mmm space group)\cite{1}. 

\begin{figure}[t]
\epsfxsize=6cm
$$
\epsfbox{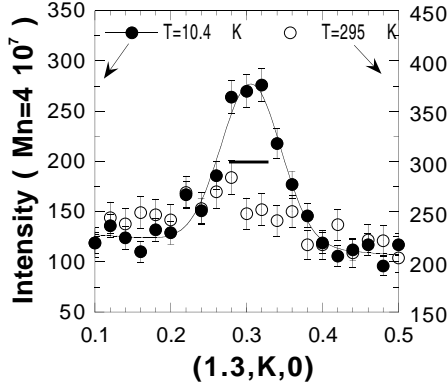}
$$
\caption{Constant-$\omega$ scans performed at $\hbar \omega$=6.2 meV around {\bf Q}=(1.3,0.3,0) along the (0,1,0) direction: T=10.4 K ($\bullet$), T=295 K ($\circ$).}
\label{fig1}
\end{figure}

Figure \ref{fig1} shows representative constant-$\omega$ scans taken in the (H,K,0)-plane: 
at $\hbar \omega$=6.2 meV and around ${\bf Q_0} $ =(1.3,0.3,0) along the (0,1,0) direction. 
The scan at 10.4 K shows a sharp maximum of intensity peaked at 
${\bf Q_0}$=(1.3,0.3,0) 
on top of a smooth background. At room temperature, this sharp peak has almost disappeared. 
The horizontal bar indicates the spectrometer resolution.

At 10.4 K, several constant-$\omega$ scans, with 6.2 meV energy transfer and performed 
along different directions ((1,0,0), (0,1,0), (1,1,0), (1,-1,0)) have revealed the
existence of comparable peaks at ${\bf Q}_0$ =${\bf q}_0$+${\bf G}$, 
where ${\bf q}_0$=($\pm$0.3,$\pm$0.3,0) $\equiv$ 
($\pm$0.6$\pi$/a,$\pm$0.6$\pi$/a,0) and 
${\bf G}$ is a zone-center  or a Z-point (001) in the (HK0)-plane. 
The fit of the data to a Gaussian profile incorporating experimental resolution function demonstrates that the peak intensity is isotropic with an 
intrinsic q-width (FWHM), $\Delta q$=0.13 $\pm$ 0.02 \AA$^{-1}$. 

\begin{figure}[tp]
\epsfxsize=6cm
$$
\epsfbox{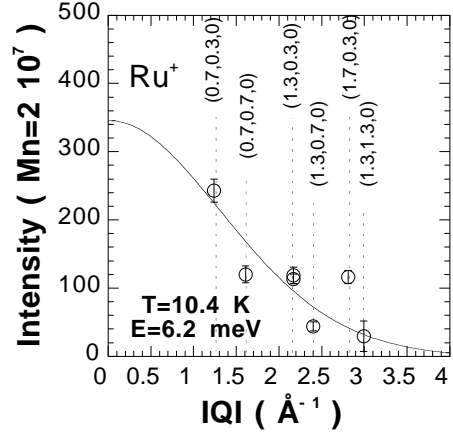}
$$
\caption{
Magnetic intensity, measured at T=10.4 K and $\hbar \omega$=6.2 meV as a function of $|Q|$. For each point, the corresponding wave vector, (H,K,L), is also reported. The full line corresponds to the square of the $Ru^+$ magnetic form factor.}
\label{fig2}
\end{figure}

The interpretation of the scattering at ${\bf q}_0$ 
as magnetic in origin 
is supported by the large number of points in reciprocal 
space where it has been observed. Further,  the lowest phonon frequencies at  ${\bf q}_0$ are above 12 meV \cite{marcus2}. In addition, in contrast 
to a phonon-related scattering that increases at large $|Q|$ or 
with temperature, the scattering at ${\bf q}_0$ decreases 
both at large wave vector (Fig.~\ref{fig2}) and at high 
temperature (Fig.~\ref{fig1}). These different points establish 
the magnetic origin of the scattering observed around ${\bf q}_0$.
In contrast, in spite of several attempts, no sizable FM spin fluctuations 
have been observed.

In a paramagnetic state, the magnetic neutron cross section per 
formula unit can be written in terms of the imaginary part of the 
dynamical spin susceptibility, $\chi"({\bf Q}, \omega)$, 
as \cite{Lovesey,xirmq};
\begin{equation}
\frac{d^2 \sigma}{d \Omega d \omega}=r_0^2 \frac{2 F^2({\bf Q})}{\pi (g  \mu_B)^2} \frac{\chi"({\bf Q},\omega)}{1-\exp(-\hbar \omega / k_BT)}
\; \label{eq:INS}
\end{equation}
where $r_0^2$=0.292 barn, $F({\bf Q})$ is the magnetic form factor and $g \simeq$2 is the 
Land\'e factor. The intensity of the scattering can be reasonably well described
by the squared magnetic form factor  of the Ru$^+$-ion \cite{formfac} (note
that the magnetic form-factor of Ru$^{4+}$ is not available) after correction
for geometrical factors related to the unfavorable shape of the sample, 
see Fig. 2. 
According to our measurements, the q-dependence of 
$\chi"$ is given by: $\chi"({\bf Q},\omega)= \chi"(q_0,\omega) 
\exp[-4\ln(2)( {\bf Q - Q_0})^2/\Delta q^2]$. 
 \begin{figure}[tp]
\epsfxsize=7cm
$$
\epsfbox{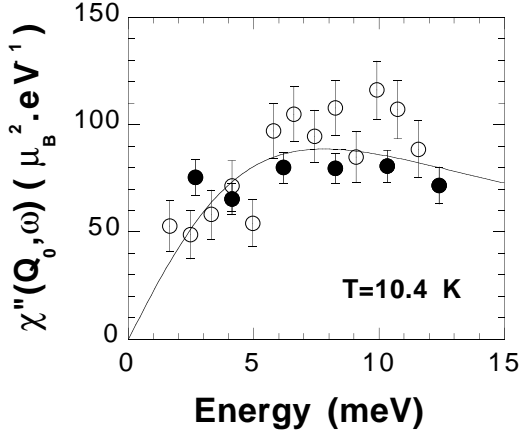}
$$
\caption{Energy dependence of the imaginary part of the dynamical magnetic susceptibility at ${\bf Q}_0$=(1.3,0.3,0) as obtained from energy scans ($\circ$) and constant energy scans around ${\bf Q}_0$ along the (0,1,0) direction ($\bullet$) (see text).}
\label{fig3}
\end{figure}

The Fermi surface in Sr$_2$RuO$_4$ is formed by three sheets \cite{9}:
one, related to the $4d_{xy}$-orbitals is quasi-2D, whereas, the two others, 
related to $4d_{xz,yz}$ orbitals are quasi-1D. 
The 1D-sheets can be schematically described by parallel planes separated 
by $\bar q$=$\pm 2\pi /3a$, running both in the x and in the y directions. 
These peculiarities give rise to dynamical nesting effects 
at the wave vectors {\bf $k$}=($\bar q,k_y$), {\bf $k$}=($k_x,\bar q$) and in particular 
at  ${\bf \bar q}$=($\bar q,\bar q$). 
The nesting effects become dominant when calculating the bare spin 
susceptibility of a non interacting metal \cite{9}, given by the Lindhard-function \cite{Lovesey}:
\begin{equation}
\chi_0 (q, \omega) = - 2 \mu_B^2 \sum_{k}\frac{f_{k+q}-f_{k}}{\varepsilon_{k+q}-\varepsilon_{k}-\hbar \omega + i \epsilon}
\; \label{eq:3}
\end{equation}
where $\epsilon \rightarrow$0, $f_k$ is the Fermi distribution function and $\varepsilon_k$ 
the quasiparticle dispersion relation. 
Our INS are in very good agreement with the predicted four spots of 
magnetic scattering situated at ${\bf \bar q}$=($\pm$2$\pi$/3a,$\pm$2$\pi$/3a) \cite{9}.
In the experiment
the incommensurate magnetic responses are actually observed slightly away, at
${\bf q}_{0 //}$=($\pm0$.6$\pi$/a,$\pm0$.6$\pi$/a), which is most
likely related to details of the band-structure  \cite{9}.

Let us now consider the energy dependence and magnitude of $\chi"(q_0,\omega)$. 
At T=10.4 K, constant-$\omega$ scans have been measured at 
{\bf Q}=(1.3,0.3,0) 
along the (0,1,0)-direction for different  transferred energies between 
2.4 and 12 meV. The magnetic response always displays a Gaussian profile, located 
at $q_0$ with an energy independent q-width, on top of a constant background. In addition, 
two energy scans have been performed at ${\bf Q}$=(1.3,0.3,0) and at {\bf Q}=(1.3,0.46,0), 
the latter providing a background reference. These measurements allow us to determine the 
energy dependence of the magnetic response at {\bf q$_0$} from 1.5 to 12 meV. 
The analysis could not be extended to higher and lower energies due to 
the contaminations by phonon\cite{marcus2} and elastic incoherent 
scattering respectively.
Using Eq.~(\ref{eq:INS}), the magnetic intensity has been converted to the dynamical 
spin susceptibility $\chi$" after correction by the thermal population factor and the 
squared magnetic  form factor reported in figure \ref{fig2}. 
We have then calibrated  $\chi$" in absolute units 
against acoustic phonons, according to a standard procedure 
\cite{phonon}. $\chi "(\omega, Q_0)$, whose energy dependence is 
reported in absolute units in Fig.\ref{fig3}, slightly increases 
up to 7 meV and then almost saturates. This energy dependence 
can be  parameterized following linear response theory:
\begin{equation}
\chi "(q_0, \omega) = \chi '(q_0, 0)  \frac{\Gamma \omega}{\omega ^2 + \Gamma ^2}
\; \label{eq:1}
\end{equation}
where $\Gamma$ is a damping energy of 9 meV and $\chi '(q_0, 0)$ = 180 $\mu_B^2.$eV$^{-1}$  corresponds to the static spin susceptibility at 
{\bf q$_0$}. It is worth emphasizing that  $\chi '(q_0, 0)$ is 
6 times larger than that at Q=0, i.e. the uniform susceptibility: 
$\tilde{\chi}=\chi '(Q=0, 0)$ = 30 $\mu_B^2.$eV$^{-1}$ 
($\simeq$ 10$^{-3}$ emu/mole) \cite{1,2,3}.
In La$_{1.86}$Sr$_{0.14}$O$_4$, usually referred to as a strongly 
correlated system, $\chi"({\bf Q}, \omega)$  at incommensurate 
wave vectors exhibits almost the same magnitude and a similar $\omega$-dependence \cite{gabe}.

\begin{figure}[t]
\vspace*{-0cm}
\epsfxsize=6.3cm
$$
\epsfbox{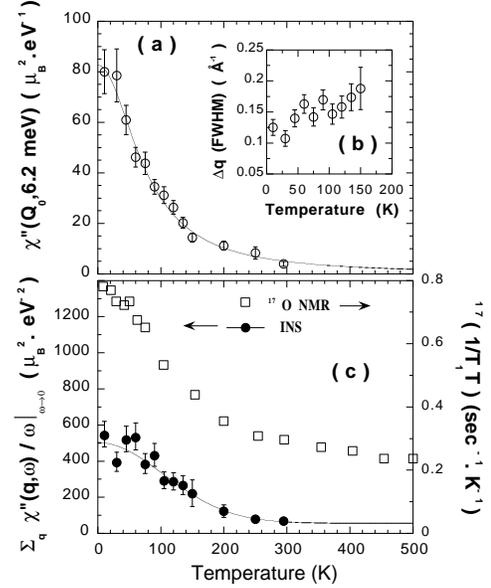}
$$
\caption[toto]{ Results from fits to a Gaussian profile of 6.2 meV constant-$\omega$ scans at 
${\bf Q}_0$=(1.3,0.3,0) along the (0,1,0): 
temperature dependences of (a)
$\chi"({\bf Q}_0,6.2 {\rm meV})$ and (b) the intrinsic q-width 
of the magnetic signal, $\Delta q$ (FWHM). (c) Comparison between $^{17}(1/T_{1}T)$ observed by $^{17}$O NMR by Imai {\it et al} \cite{12} ($\Box$) 
 and the incommensurate contribution 
calculated from our INS measurements ($\bullet$).
Assuming $\Lambda$=33 kOe/$\mu_B$\cite{Aq2}
the two scales in this figure are identical.
}
\label{fig4}
\end{figure}

In Sr$_2$RuO$_4$, electronic correlations are incorporated in  RPA calculations: the spin 
susceptibility  $\chi (q, \omega)$ becomes enhanced through the Stoner-factor 
$I(q)$\cite{7,9}:
\begin{equation}
\chi (q, \omega) = \frac{\chi_0(q, \omega)}{1 - \frac{I(q)}{2 (\mu_B)^2} \chi_0(q, \omega)}
\; \label{eq:2}
\end{equation}
The q-dependence of the Stoner factor, for an individual RuO$_2$ plane, 
reflects the fact 
that FM interactions are favored over antiferromagnetic interactions in Sr$_2$RuO$_4$ :
in our units, $I(q)= 0.86/(1+0.8(a/\pi)^2q^2)$ eV (q in \AA$^{-1}$)\cite{7,9}. INS results point 
towards a strong enhancement of the spin susceptibility by the Stoner factor 
(see Eq.~(\ref{eq:2})), such that the system should be close to a magnetic instability at 
${\bf q}_0$. With $\chi '(q_0, 0)$ = 180 $\mu_B^2.$eV$^{-1}$, one deduces from  Eq.~(\ref{eq:2}) that $\frac{I(q_0)}{2(\mu_B)^2}\chi_0(q_0, 0) \simeq$0.99, instead of being larger than 1 for a magnetic instability. Thus, incommensurate spin fluctuations are stronger than FM 
fluctuations in Sr$_2$RuO$_4$, as suggested in ref.~\cite{9}.

The temperature dependence of both $\chi "({\bf Q}_0, 6.2 {\rm meV})$ and the intrinsic q-width 
are reported in Fig.~\ref{fig4}, as deduced from constant-$\omega$ scans performed at 6.2 meV 
around {\bf Q}=(1.3,0.3,0) along the (0,1,0) direction at 
different temperatures. 
$\chi "({\bf q}_0, 6.2 {\rm meV})$ exhibits a sharp decrease upon temperature increase and 
simultaneously the magnetic response broadens (the width of the signal can be reliably 
determined only up to 200 K). 
The T-dependence of $\chi"$ observed in INS measurements may be described
by the out-smearing of the Fermi surface due to thermal hopping of electrons 
into unoccupied states (see the numerator in Eq.~(\ref{eq:3})),
yielding a lowering of the dynamical susceptibility at ${\bf q}_0$
and its broadening in q-space. The T-dependence of the magnetic response at ${\bf q}_0$ 
can indeed be qualitatively reproduced \cite{pfeuty} using Eqs. (\ref{eq:3})-(\ref{eq:2}) 
and a description of the LDA band structure by three mutually non-hybridizing tight-binding 
bands \cite{7}.


INS measurements point out the existence of strong magnetic response at ${\bf q}_0$, 
but do not reveal any sizeable FM fluctuations. 
In contrast, the uniform spin susceptibility 
\cite{1,2,3} and the Knight shift measurements \cite{12,13} provide 
evidence of strong FM correlation in Sr$_2$RuO$_4$. 
However, the delicate balance between 
FM and incommensurate spin fluctuations should 
become visible in the spin-lattice 
relaxation rate $T_1$ measured by both $^{17}$O and  $^{101}$Ru 
NMR experiments\cite{12,13}. 
These NMR-techniques probe the low energy spin fluctuations
($\omega \rightarrow 0$ with respect to INS measurements); furthermore,
they  integrate the fluctuations in q-space.
Since the INS studies have determined the incommensurate fluctuations
on an absolute scale we may estimate their contribution
to $(1/T_1T)$, $^{INS} (1/T_1T)$. 
In general $(1/T_1T)$ probes the q-summation of the 
the imaginary part of the susceptibility divided by the 
frequency in the limit $\omega \rightarrow 0$ (i.e. its initial slope),
${\sum_{q} \frac{\chi "(q,\omega)}{ \omega} }\vert_{\omega 
\rightarrow 0}$; its temperature dependency  
is shown in Fig. 4.c (left scale).

What renders the quantitative comparison between the INS and NMR-results 
more difficult is the estimate of the hyperfine field whose 
q-dependent Fourier transform, $A(q)$,  weights the susceptibility in 
NMR-studies. Considering that INS magnetic fluctuations are sharply 
peaked around {\bf q$_0$}, one may approximate $A(q)$=$A(q_0)$, and 
gets\cite{berthier},

\begin{equation} 
^{ INS}(1/T_1T) \simeq \frac{k_B\gamma_n^2}{(g\mu_B)^2} 
 \left. | A(q_0) |^2 {\sum_{q}  \frac{\chi "(q,\omega)}{ \omega} }\right\vert_{\omega \rightarrow 0}
\; \label{eq:1/T1}
\end{equation}
with $| A(q_0) |^2= \Lambda ^2 [1+1/2(\cos(2\pi 0.3)+\cos(2\pi / a0.3))]$
($\Lambda$ = 33 kOe/${\mu _B}$\cite{Aq2}) for $^{17}$O and 
$A(q_0)$=-299 kOe/$\mu_B$\cite{10,12} for $^{101}$Ru. 
Using these values,  we directly compare $^{INS}(1/T_1T)$ 
with the measured $^{17}$O $(1/T_1T)$  in Fig. \ref{fig4}.c (right scale). 
Clearly,  the spin fluctuations at ${\bf q}_0$ significantly
contribute to $^{17}(1/T_1T)$, and can explain a large part of 
the reported T-dependence\cite{12,13}. Similar calculation for the
$^{101}$Ru $(1/T_1T)$ (not shown) yields even a stronger contribution.
The remaining parts in  $^{17,101}(1/T_1T)$, likely 
associated with FM excitations, should exhibit a less pronounced  
T-dependence similar to that of the uniform static spin 
susceptibility. 
Furthermore, assuming a weak q-dependence for these FM 
excitations\cite{pfeuty}, the comparison of NMR and INS measurements 
allows us to estimate the ferromagnetic characteristic energy
to be of the order of 50 meV.
This rather elevated value actually  provides a satisfactory explanation 
for the absence of FM fluctuations in our INS measurements.

To conclude, our INS measurements demonstrate the existence of incommensurate 
spin fluctuations  related to dynamical nesting properties 
of the Sr$_2$RuO$_4$ Fermi surface. 
Our data suggest that the system is close to a magnetic instability at 
${\bf q}_{0//}$=($\pm 0.6 \pi /a$,$\pm 0.6 \pi /a$). 
The comparison of INS and $^{17}(1/T_1T)$ measurements  suggests 
that the FM fluctuations are transferred to higher energy with respect to 
the spin fluctuations at ${\bf q}_0$. 
All these results cast some doubt on the predominant role of FM spin 
fluctuations in  the mechanism of superconductivity in Sr$_2$RuO$_4$. 


We wish to acknowledge P. Pfeuty, J. Bobroff and Ph. Mendels 
for helpful discussions.


\begin{references}

\bibitem{1} Y. Maeno {\it et al.}, Nature {\bf372}, 532 (1994).

\bibitem{2} Y. Maeno {\it et al.},J. Phys. Soc. Jpn. {\bf66}, 1405 (1997).

\bibitem{3} J. J. Neumeier {\it et al.}, Phys. Rev. B {\bf50}, 17910 (1994).

\bibitem{4} T. M. Rice {\it et al.}, J. Phys. Cond. Matt. {\bf7}, L643 (1995).

\bibitem{5} I. I. Mazin {\it et al.}, Phys. Rev. B {\bf56}, 2556 (1997).

\bibitem{6} K. Machida {\it et al.}, J. Phys. Soc. Jpn. {\bf65}, 3720 (1996).

\bibitem{oguchi} T. Oguchi, Phys. Rev. B {\bf 51}, 1385 (1995).

\bibitem{7} I. I. Mazin {\it et al.}, Phys. Rev. Lett {\bf79}, 733 (1997).

\bibitem{8} A. P. Mackenzie {\it et al.}, Phys. Rev. Lett. {\bf76}, 3786  (1996).

\bibitem{9} I. I. Mazin and D.J. Singh, Cond-mat{\bf/9902193}.

\bibitem{10} K. Ishida {\it et al.}, Phys. Rev. B {\bf 56}, R505 (1997).

\bibitem{11} G. M. Luke et al, Nature {\bf 394}, 558 (1998).

\bibitem{ishida} K. Ishida et al, Nature {\bf396}, 658 (1998).

\bibitem{12} T. Imai et al, Phys. Rev. Lett. {\bf 81}, 3006  (1998).

\bibitem{caruo} M. Braden {\it et al.},  Phys. Rev. B {\bf 58}, 847 (1998).

\bibitem{nakatsuji}S. Nakastuji {\it et al.}, J. Phys. Soc. Jpn. {\bf 66}, 1868 (1997).


\bibitem{marcus2} M. Braden {\it et al.},  Phys. Rev. B {\bf 57}, 1236  (1998).

\bibitem{formfac} {\it International Tables of Crystallography}, Vol. C, eds A. J. C. Wilson, Kluwer Academic publisher (1995).

\bibitem{Lovesey} S. W. Lovesey, {\it Theory of Neutron Scattering from Condensed matter}. (Clarendon, Oxford,1984), Vol.2. the kinematic factor $k_F/k_I$ has been omitted in Eq. (1) for sake of simplicity.

\bibitem{xirmq} The spin susceptibility is associated with fluctuations of a single spin component: $\chi({\bf Q}, \omega)=\chi^{\alpha \beta}= -(g \mu_B)^2 \frac{i}{\hbar} \int_{0}^{\infty} dt \exp(i \omega t) \langle [S_i^{\alpha}(t),S_j^{\beta}] \rangle$, where $\alpha, \beta$ are Cartesian coordinates. $\chi$ is half of the transverse spin susceptibility $\chi^{+-}$.

\bibitem{phonon} We have used acoustic phonons in scans (2+$\xi$,0,0) and (2,$\xi$,0)
for $\hbar \omega$= 6.75 and 8.26 meV (sound velocity $\simeq$ 20.8 and 43.9 meV.\AA, respectively)\cite{marcus2}.

\bibitem{gabe} G. Aeppli {\it et al.}, Science, {\bf 278}, 1432  (1997).

\bibitem{pfeuty} P. Pfeuty, private communication.

\bibitem{13} H. Mukuda {\it et al.}, J. Phys. Soc. Jpn. {\bf 67}, 3945 (1998).

\bibitem{berthier} C. Berthier {\it et al.}, J. Phys. I France {\bf 6}, 2205 (1996); R.E. Walstedt {\it et al.}, Phys. Rev. Lett. {\bf 72}, 3610  (1994).

\bibitem{Aq2} For a magnetic field applied along the c axis, $\Lambda^2$=$|A_{//}|^2+|A_{\perp}|^2$, with $A_{//}$=-18.5 kOe/$\mu_B$ and  $A_{\perp}$=-26.8 kOe/$\mu_B$\cite{13}. 


\end{references}
\end{document}